\documentclass[twocolumn, prl, superscriptaddress]{revtex4-1}
\usepackage{amssymb}
\usepackage{amsmath}
\usepackage{epsfig}
\usepackage{color}
\usepackage{graphics, graphicx}
\usepackage{bbold}
\usepackage{psfrag}
\usepackage{mathcomp}
\usepackage{subfigure}
\usepackage{verbatim}
\usepackage{color}
\usepackage{ulem}
\usepackage[colorlinks,citecolor=blue]{hyperref}

\begin{document}
\title{Tunable non-reciprocal quantum transport through a dissipative Aharonov-Bohm ring in ultracold atoms}
\author{Wei Gou}
\thanks{These authors contributed equally to this work}
\author{Tao Chen}
\thanks{These authors contributed equally to this work}
\author{Dizhou Xie}
\author{Teng Xiao}
\affiliation{%
Interdisciplinary Center of Quantum Information, State Key Laboratory of Modern Optical Instrumentation, and Zhejiang Province Key Laboratory of Quantum Technology and Device of Physics Department, Zhejiang University, Hangzhou 310027, China
}%
\author{Tian-Shu Deng}
\affiliation{Institute for Advanced Study, Tsinghua University, Beijing, 100084, China}
\author{Bryce Gadway}
\affiliation{Department of Physics, University of Illinois at Urbana-Champaign, Urbana, IL 61801-3080, USA}
\author{Wei Yi}
\email{wyiz@ustc.edu.cn}
\affiliation{CAS Key Laboratory of Quantum Information, University of Science and Technology of China, Hefei 230026, China}
\affiliation{CAS Center For Excellence in Quantum Information and Quantum Physics, Hefei 230026, China}
\author{Bo Yan}
\email{yanbohang@zju.edu.cn}
\affiliation{%
Interdisciplinary Center of Quantum Information, State Key Laboratory of Modern Optical Instrumentation, and Zhejiang Province Key Laboratory of Quantum Technology and Device of Physics Department, Zhejiang University, Hangzhou 310027, China
}%
\affiliation{%
 Collaborative Innovation Centre of Advanced Microstructures, Nanjing University, Nanjing, 210093, China
}%
\affiliation{%
 Key Laboratory of Quantum Optics, Chinese Academy of Sciences, Shanghai, 200800, China
}

\date{\today}

\begin{abstract}
We report the experimental observation of tunable, non-reciprocal quantum transport of a Bose-Einstein condensate in a momentum lattice.
By implementing a dissipative Aharonov-Bohm (AB) ring in momentum space and sending atoms through it, we demonstrate a directional atom flow by measuring the momentum distribution of the condensate at different times. While the dissipative AB ring is characterized by the synthetic magnetic flux through the ring and the laser-induced loss on it, both the propagation direction and transport rate of the atom flow sensitively depend on these highly tunable parameters. We demonstrate that the non-reciprocity originates from the interplay of the synthetic magnetic flux and the laser-induced loss, which simultaneously breaks the inversion and the time-reversal symmetries.
Our results open up the avenue for investigating non-reciprocal dynamics in cold atoms, and highlight the dissipative AB ring as a flexible building element for applications in quantum simulation and quantum information.
\end{abstract}
\maketitle

Quantum transport, a fundamental property and a key probe of quantum many-body systems, lies at the core of seminal discoveries in condensed-matter physics such as superconductivity~\cite{SCreview} and topological materials~\cite{TPreview1,TPreview2}. Besides reciprocal transport, where the transfer function of energy or particle between two points in space is symmetric in the direction of flow, non-reciprocal transport, both quantum mechanical and classical, also exists in various physical contexts, and finds applications in electric diodes, non-reciprocal optical, optomechanical device~\cite{Sounas2017,nonopt1, nonopt2, nonopt3, nonopt4, nonopt5,nonopt6,nonrecph,Wang2013,nonrecWG,StefanoOL,Jin2018}, and non-linear metamaterials~\cite{nonrecMeta,Feng2013}. Understanding and controlling non-reciprocal quantum transport is of fundamental importance for the study of many-body dynamics~\cite{Duca2015,Seif2018}, the quantum simulation of exotic models~\cite{Tomoki2019,Eduardo2019}, and the design of useful quantum device for quantum information~\cite{Stannigel2012,Kim2015,Hurst2018,Shabir2018,Li2018}.

With flexible controls and versatile detection schemes, quantum gases have proved to be an important physical platform for the study of quantum transport~\cite{Billy2008,Kondov2011,Schreiber2015, Clark2017, Feng2019, Meier2016}. For example, superfluidity in multiple-connected geometries has been investigated using a Bose-Einstein condensate (BEC) in ring traps ~\cite{Eckel2014,Eckel2014a}, and quantized conductance has been reported for the transport of cold atoms through a point contact~\cite{Chien2015,Haeusler2017}. However, non-reciprocal quantum transport has yet to be implemented with cold atoms, where the interplay of non-reciprocity and the highly tunable parameters of the many-body system {\it per se} holds fascinating potentials for both quantum simulation and quantum information~\cite{Francois2019,Xu2019}.

\begin{figure*}[tbp]
\includegraphics[width=0.8\textwidth]{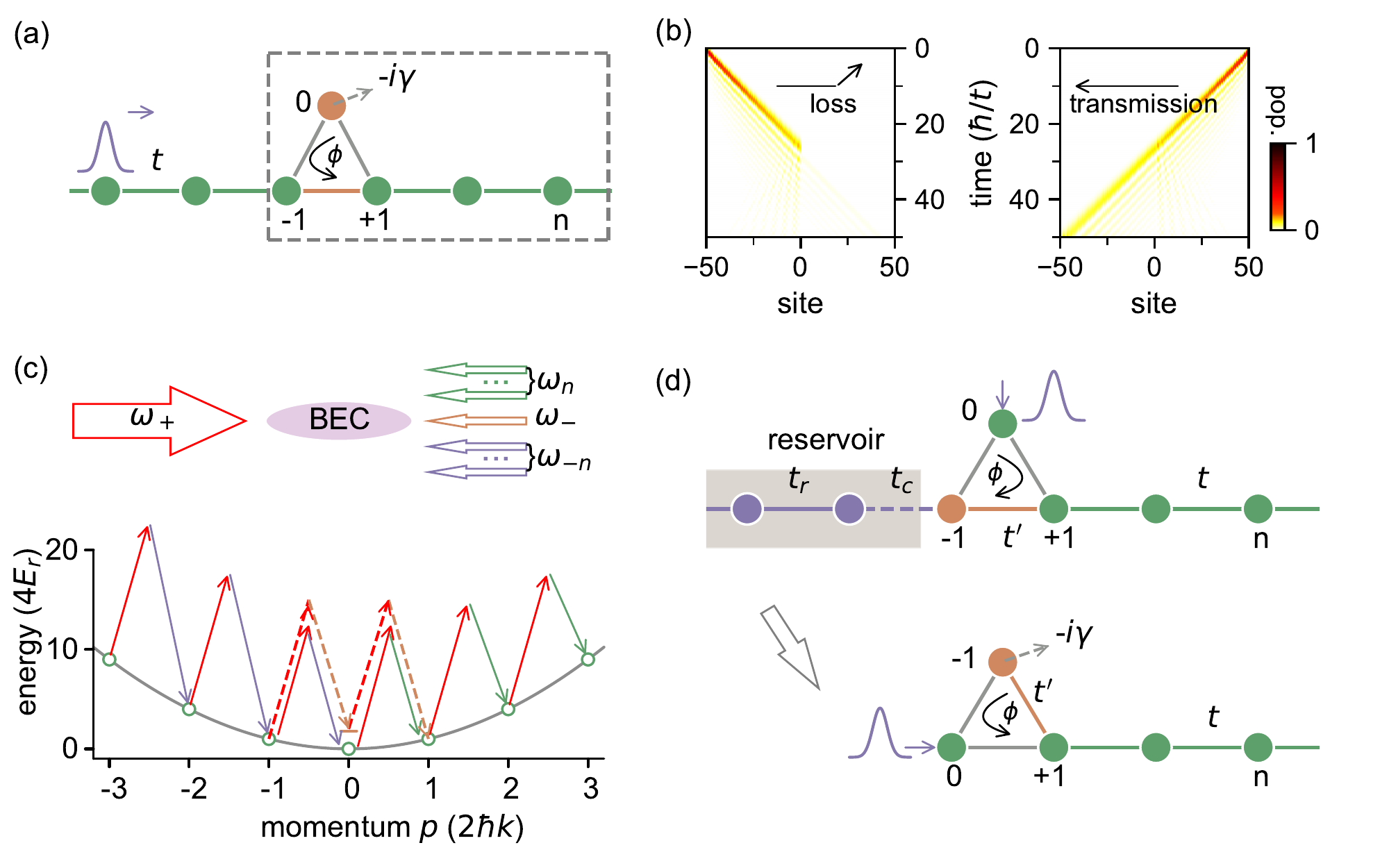}
\caption{\label{fig1} (a) Schematic illustration of a dissipative AB ring (the $3$-site triangle within the dashed box) along a lattice. {The green, grey, and orange bonds correspond to hopping rates $t$, $t e^{-i\phi/2}$, and $t'$, respectively, between adjacent sites.}
The vertex of the ring (site $0$ in orange) features an on-site loss with rate $\gamma$.
(b) Numerical simulation of non-reciprocal atom transport through the AB ring in (a), where we initialize atoms on the lattice sites $\pm 50$ to the left or right of the ring, and plot the time-dependent population distribution~\cite{sm}. We fix $\gamma / t = 1$ and $\phi=3\pi/2$ for the calculation, in which case atoms incoming from the left are lost to the reservoir through site $0$, whereas atoms incoming from the right can transport through.
(c) Illustration of the experimental implementation of the momentum lattice and the AB ring. The momentum lattice is formed with multiple pairs of two-photon Bragg transitions (solid arrows), while momentum states $|n=1\rangle$ is coupled to $|n=-1\rangle$ with a four-photon second-order Bragg process (dashed arrows). Here the frequency components $\omega_n$ of the left-going Bragg laser satisfies $\omega_n = \omega_+ - 4(2n+1)E_r/\hbar$, while $\omega_- = \omega_+$.
(d) Implementation of on-site loss and the mapping of the system to (a). The left side of the lattice with $n<-1$ is mapped to a reservoir, with hopping rate $t_r$ within the reservoir and $t_c$ between the system and the reservoir. The laser-induced hopping rate between momentum states $|-1\rangle$ and $|1\rangle$ is $t'$, which allows us to map $|-1\rangle$ to lattice site $0$ in (a) with an effective loss rate $\gamma\approx t_c^2/t_r$.
}
\end{figure*}

In this work, we report the experimental observation of non-reciprocal quantum transport of a BEC through a dissipative Aharonov-Bohm (AB) ring. Coupling discrete momentum states using multi-frequency Bragg lasers, we implement a dissipative AB ring on a momentum lattice, where both the synthetic magnetic flux through the ring and the laser-induced loss on the ring are easily tunable.
By measuring the atomic momentum distribution at different times, we experimentally probe the transport of atoms through the AB ring, and observe a parameter-dependent, directional atom flow, which originates from the interplay of the synthetic magnetic flux and the laser-induced loss.

\begin{figure}[tbp]
\includegraphics[width=0.48\textwidth]{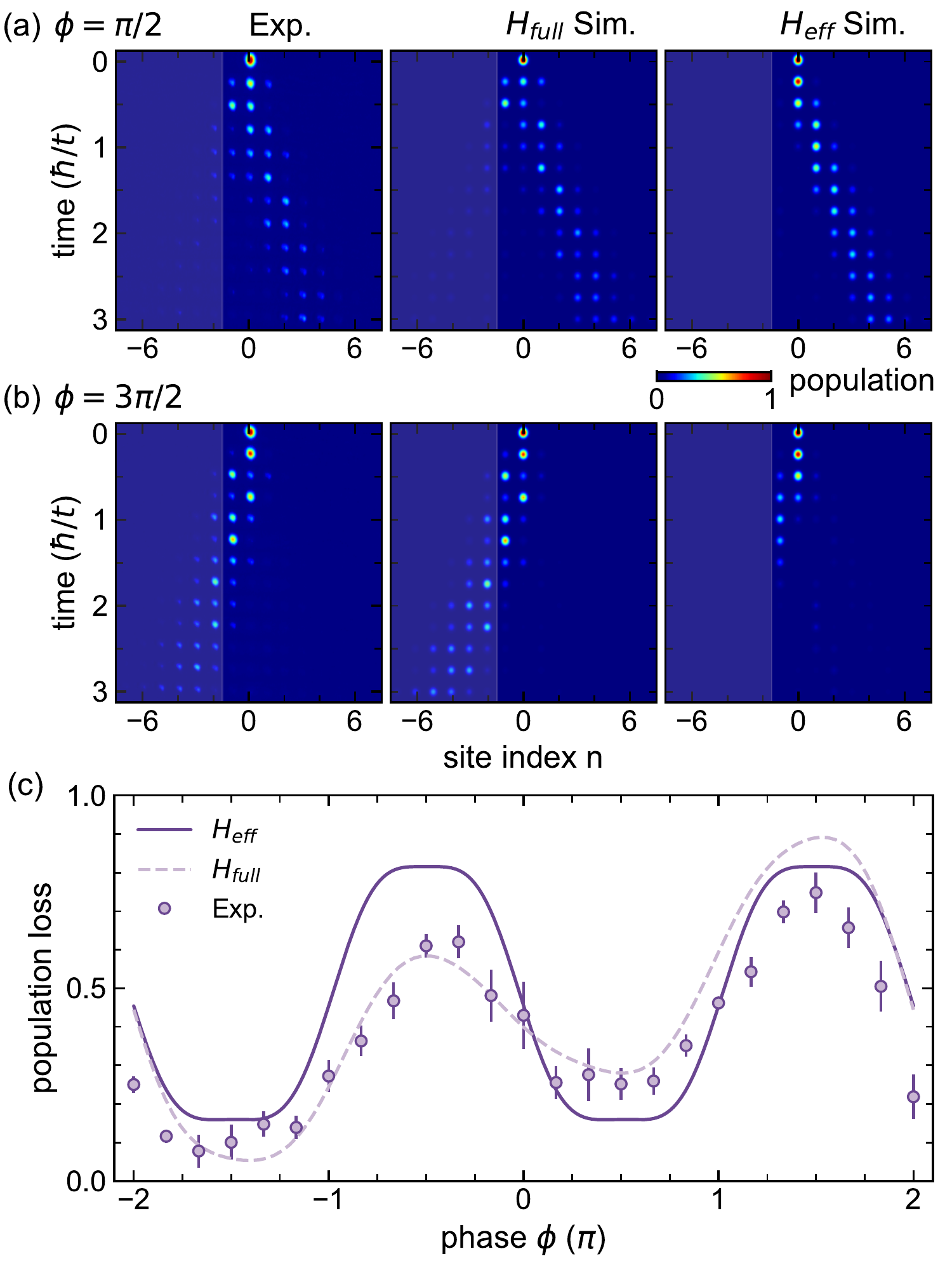}
\caption{\label{fig2}(a)(b) Phase-dependent transport for (a) $\phi=\pi/2$, and (b) $\phi=3\pi/2$. From left to right, the panels are experimental data, numerical simulation using the full Hamiltonian, and simulation using the effective Hamiltonian~\cite{sm}. The masked regions correspond to the reservoir. (c) The population loss $\mathcal{P}_\ell$ as a function of $\phi$, after an evolution time of $3\hbar/t$. Experimental data are shown as purple dots with error bars. The dashed line is from the numerical simulation with the full Hamiltonian, and the solid line is the prediction from the effective Hamiltonian (\ref{Eq2}). For all cases, we fix the effective loss rate $\gamma/t=1$, and initialize the BEC on site $|0\rangle$.}
\end{figure}

{\it Implementing dissipative AB ring:---}
As illustrated in Fig.~\ref{fig1}(a), the key element for non-reciprocal transport, the dissipative AB ring, consists of a closed triangle with three lattice sites. The synthetic magnetic flux through the ring is generated by the phases of the hopping rates within the triangle. The on-site loss with a rate $\gamma$ on the vertex of the triangle makes the AB ring dissipative. To demonstrate the non-reciprocal transport through the ring, we perform numerical simulations on the time evolution of a single-particle initialized to the left or right of the ring on the lattice~\cite{sm}. As shown in Fig.~\ref{fig1}(b), atom transport can be made unidirectional under the parameters $\gamma / t = 1$ and $\phi=3\pi/2$, i.e., they are only allowed to propagate through the ring when injected from the left. Note that directional transport from the right to the left is also possible by choosing different parameters $\gamma$ and $\phi$.

Our experimental implementation of the dissipative AB ring is schematically illustrated in Fig.~\ref{fig1}(c). Starting with a BEC of $6\times 10^4$ $^{87}$Rb atoms in a weak crossed-dipole trap with trapping frequencies $2\pi\times(115,40,100)$Hz \cite{Xie2018}, we create a one-dimensional momentum lattice along the $y$-direction, using a series of two-photon Bragg transitions to couple discrete momentum states $|n\rangle$ ($n\in \mathbb Z$)~\cite{Meier2016,An2017, Xie2019}.
The Bragg transitions are driven by counter-propagating, far-detuned laser pairs with the wavelength $\lambda=1064~\text{nm}$, whose multi-frequency components (with frequencies $\omega_n$) are generated by acoustic optical modulators (AOMs).
The resulting discrete momentum state $|n\rangle$ along the lattice has the momentum $p=2n\hbar k$, where the wave vector of the lasers $k=2\pi/\lambda$, and the single-photon recoil energy $E_r=(\hbar k)^2/2m=h\times2.03$kHz, with $m$ the atomic mass.
As the nearest-neighbour sites are coupled by resonant Bragg transitions, the hopping rate $t$ is determined by the effective Rabi frequency, and we fix it at $t=h\times 1.25(2)$kHz throughout our experiment. Under these conditions, the interaction effects on the dynamics is negligible~\cite{sm}.

To form a closed triangular AB ring, we couple the momentum states $|n=-1\rangle$ and $|n=1\rangle$ using a four-photon process, with the effective coupling rate $t'=h\times 1.26(4)$kHz. The four-photon process is induced by a pair of lasers with frequencies $\omega_\pm$ ($\omega_+=\omega_-$), as shown in Fig.~\ref{fig1}(c)~\cite{Giese2013, An2018}. Denoting the relative phases of the frequency components $\{\omega_{-},\omega_{0,-1}\}$, with respect to $\omega_+$, as $\{\phi_{-},\phi_{0,-1}\}$, we implement the synthetic magnetic flux $\phi$ of the AB ring by setting $2\phi_--\phi_{-1}-\phi_0=\phi$. Here we choose a symmetric setting with $\phi_- = 0$ and $\phi_0 = \phi_{-1} = -\phi/2$. Furthermore, we introduce detunings for different frequency components, to compensate for the site-dependent Stark shifts induced by off-resonant Bragg transitions~\cite{sm}.

To implement the on-site loss, we use the left side of the lattice ($n<-1$) as a reservoir.
As shown in Fig.~\ref{fig1}(d), the laser-induced hopping rate is $t_r=t$ within the reservoir, whereas the hopping rate between sites $|-1\rangle$ and $|-2\rangle$ is $t_c$. {The loss mechanism is best understood in the extreme limit of $t_c\ll t_r$, where it takes a long time for a coherent population exchange between the system ($n\geq -1$) and the reservoir ($n<-1$). Any population of the reservoir (sites with $n<-1$) is therefore considered as loss during this period.
It follows from the second-order perturbation that the effective on-site loss rate for $|-1\rangle$ at short times is $\gamma\approx t_c^2/t_r^{}$~\cite{Lapp2019}. For larger $t_c$, $\gamma$ should deviate from $t_c^2/t_r$. However, as we show later, such a deviation does not lead to significant error up to $\gamma\sim t$.} Finally, in the experiment, all the parameters $t$, $t'$, $t_c$ and $t_r$ can be independently tuned by adjusting the laser intensities for the corresponding frequency component~\cite{sm}.

With these, the effective Hamiltonian of the system is written as
\begin{align}\label{Eq2}
H_{\rm eff} &= -i\gamma c_{-1}^\dagger c_{-1}^{} + \Big[\sum_{n\geq1} t c_n^\dagger c_{n+1} + te^{-i\frac{\phi}{2}}c_{-1}^\dag c_0\nonumber\\
&+ te^{-i\frac{\phi}{2}}c_0^\dag c_1 + t'c_{1}^\dagger c_{-1}^{} + \text{H.c.}\Big],
\end{align}
where $c_n$ ($c_n^\dag$) is the annihilation (creation) operator on site $n$ of the momentum lattice. Note that we neglect the higher-order, off-resonant couplings, which can be taken into account by considering a time-dependent full Hamiltonian. The Stark shifts of these off-resonant couplings, however, are compensated by shifting the frequencies, as we discuss earlier~\cite{sm}.
Hamiltonian (\ref{Eq2}) describes a dissipative AB ring [see Fig.~\ref{fig1}(d)], where all parameters are highly tunable experimentally.

{\it Tunable non-reciprocal transport:---}
We first investigate the special case with an effective loss rate $\gamma=t$ for a $15$-site chain, with the BEC naturally initialized on the momentum-lattice site $|0\rangle$. We show the measured momentum-space density distribution at different times of the evolution in Figs.~\ref{fig2}(a) ($\phi = \pi/2$) and \ref{fig2}(b) ($\phi=3\pi/2$), where the reservoir region is masked. The experimental data fit well with both numerical simulations using the full and the effective Hamiltonians, and the dynamics on the momentum lattice is quite different for different values of $\phi$.
When $\phi = \pi/2$, transmission dominates: BEC atoms pass through the dissipative AB ring to populate lattice sites with $n>1$ [Fig.~\ref{fig2}(a)]. In contrast, when $\phi = 3\pi/2$, loss dominates: BEC atoms are blocked by the ring, leaving lattice sites with $n>1$ mostly unpopulated. Our observation is therefore consistent with the directional transport illustrated in Fig.~\ref{fig1}(b).

Next, we experimentally characterize transport properties of the ring by tuning the flux parameter $\phi$. To quantitatively analyze the phase-dependent transport, we introduce the population loss $\mathcal{P}_\ell$, defined as the total population in the reservoir
\begin{equation}
\mathcal{P}_\ell=\sum_{n<-1}\rho_n,
\end{equation}
where $\rho_n$ is the population of the momentum state $|n\rangle$.
By definition, $\mathcal{P}_\ell$ measures the atom population lost to the reservoir: transmission dominates when $\mathcal{P}_\ell$ is small; and loss dominates when $\mathcal{P}_\ell$ is large. For our experiment, we measure $\mathcal{P}_\ell$ after an evolution time of $\tau = 3\hbar/t~(\sim 384~\mu\text{s})$ for different values of the flux parameter $\phi$. As shown in Fig.~\ref{fig2}(c), $\mathcal{P}_\ell$ oscillates with varying $\phi$, reflecting a sensitive dependence of the transport ability of the dissipative AB ring on the flux.
{We compare the experimental data with simulations under both the effective Hamiltonian (solid line) and the full Hamiltonian (dashed line). Apparently, the measurements agree qualitatively well with numerical simulations, with the agreement quantitatively better for the full-Hamiltonian simulation, due to the inclusion of higher-order, off-resonance processes. We note that these off-resonant couplings are also responsible for the difference in $\mathcal{P}_\ell$ at $\phi=0,\pm2\pi$, calculated using the full Hamiltonian.}
Importantly, the measured $\mathcal{P}_\ell$ suggests a transmission-dominant behavior at $\phi=\pi/2$, and a loss-dominant behavior at $\phi=3\pi/2$. We note that atoms passing through the AB ring to the right with the flux parameter $\phi$ and atoms passing to the left with $2\pi-\phi$ are reverse processes, since their corresponding terms in Hamiltonian (\ref{Eq2}) are Hermitian conjugate to each other up to a gauge transformation. Hence, our experimental observation clearly demonstrates that the transport through the dissipative AB ring is non-reciprocal.  As we show below, such a non-reciprocity originates from the interplay of the synthetic flux and the on-site loss, which simultaneously break inversion and time-reversal symmetries.

\begin{figure}[tbp]
\includegraphics[width=0.48\textwidth]{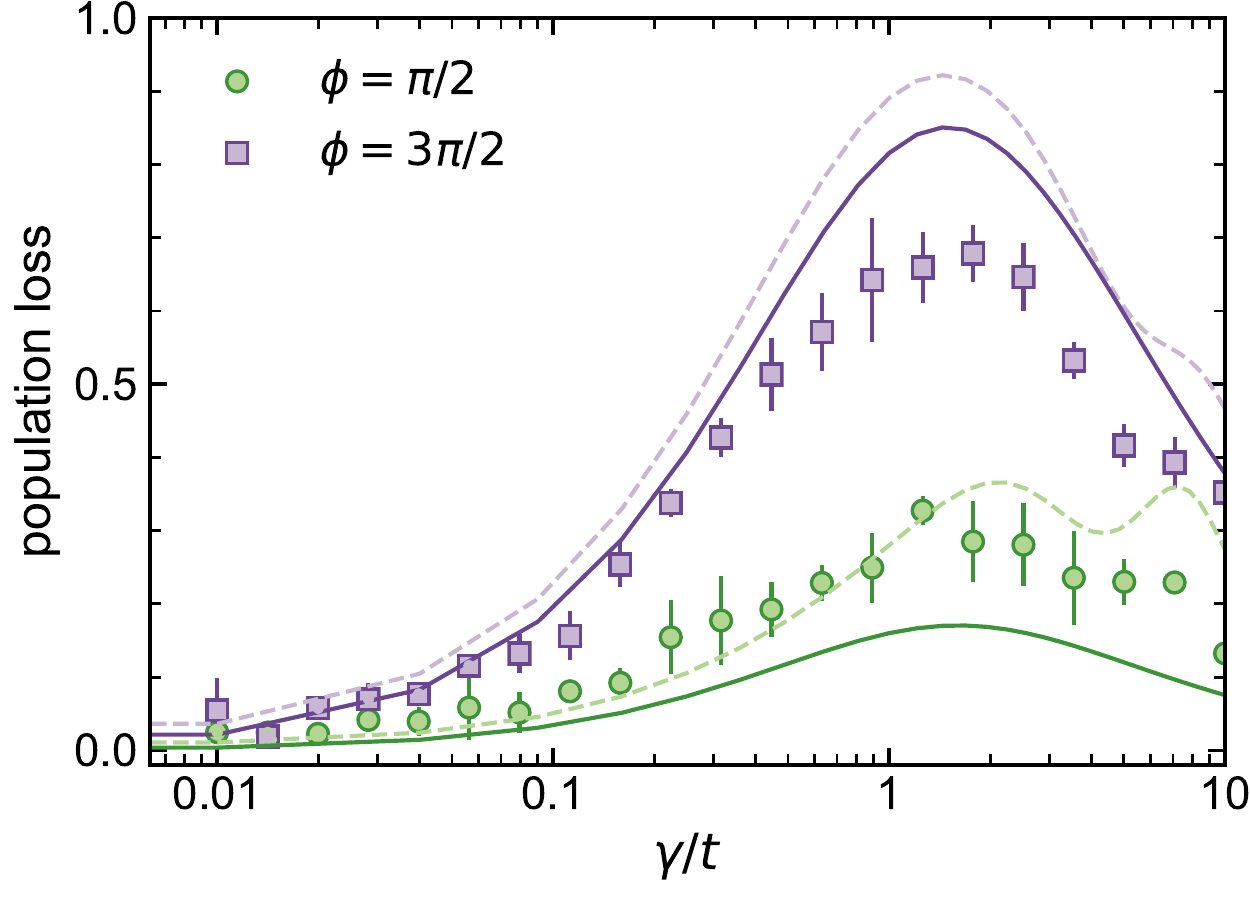}
\caption{\label{fig3} Dependence of the population loss $\mathcal{P}_\ell$ on the effective loss rate $\gamma$. Two data sets are shown, with the synthetic magnetic flux $\phi = \pi/2$ (green circles) and $3\pi/2$ (purple squares), respectively. We also show results from numerical simulations using the effective (solid lines) and the full Hamiltonians (dashed lines). For all cases, the evolution time is taken to be $\tau=3\hbar/t$.
}
\end{figure}

{\it Impact of the on-site loss:---}
We now explore the dependence of the atom transport on the on-site loss rate $\gamma$. With a fixed hopping rates $t_r = t $ and by tuning $t_c$ from $0.01t$ to $4t$ in the experiment, we are able to change $\gamma/t$ by $3$ orders of magnitude. Similar to the previous experiment, we use $\mathcal{P}_\ell$ to characterize the transport. Figure~\ref{fig3} shows two sets of experimental measurements with $\phi=\pi/2$ and $\phi=3\pi/2$, respectively.
For both measurements, $\mathcal{P}_\ell$ peaks at an intermediate $\gamma/t\sim 1$, where it is most sensitive to the flux parameter $\phi$, as discussed above.
Further, our experimental data fit better with numerical simulations for $\gamma<t$, which is easy to understand, since the expression $\gamma\approx t_c^2/t_r^{}$ is no longer a reasonable approximation for $\gamma>t$.

Physically, the loss dependence of the transport can be understood as follows. In the weak-coupling regime with $t_c \to 0$, site $|-1\rangle$ is effectively disconnected from the reservoir. On the other hand, in the strong coupling regime, quantum Zeno effect suppresses the population loss~\cite{Itano1990}.
It follows that, the non-reciprocal transport can only occur when $\gamma$ is neither too small nor too large. Our experiments thus
reveals that the interplay of the synthetic magnetic flux and the on-site loss gives rise to the non-reciprocal transport of the dissipative AB ring.

\begin{figure}[tbp]
\includegraphics[width=0.48\textwidth]{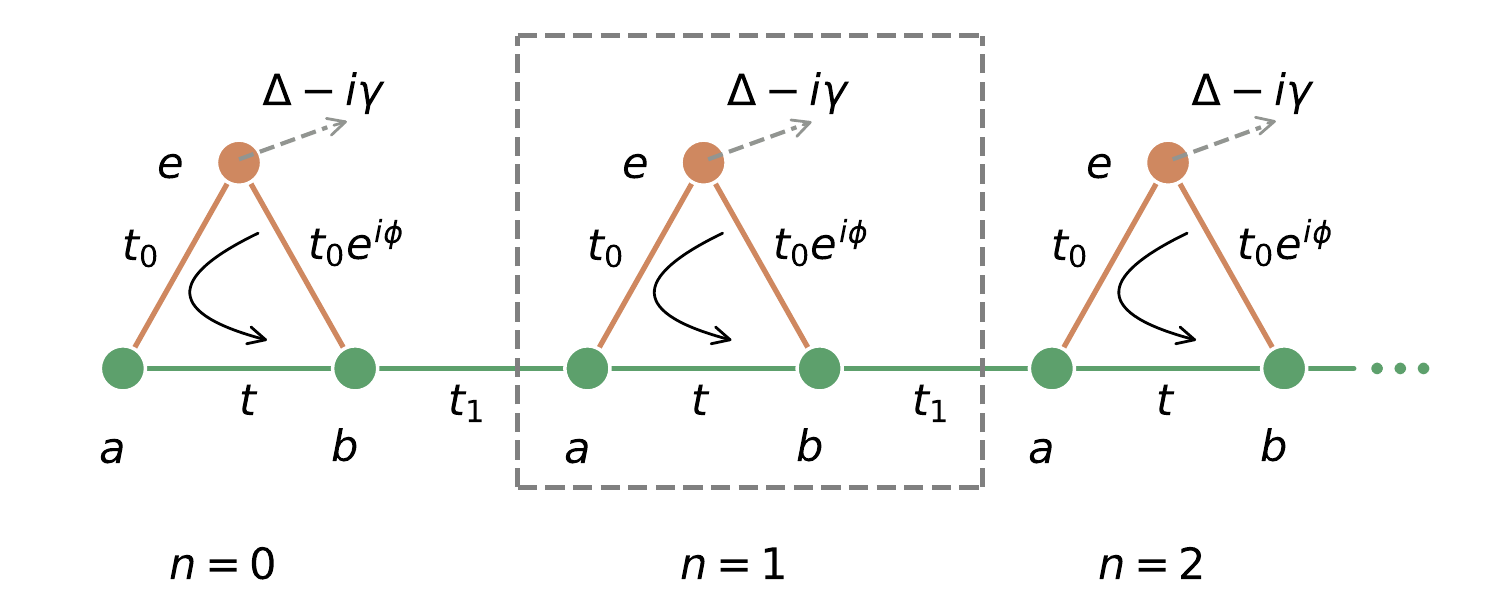}
\caption{\label{fig4} Schematic illustration of a lattice with coupled dissipative AB rings.}
\end{figure}

{\it Discussions and outlook:---}
The dissipative AB rings demonstrated here can find useful applications in quantum simulation and quantum information.
As a concrete example and outlook, we now discuss in more detail the possibility of simulating highly non-trivial non-Hermitian topological models using dissipative AB rings.

We consider a series of coupled dissipative AB rings, with hopping rates $t_0$, $t$ and $t_1$ as indicated in Fig.~\ref{fig4}. Besides the on-site loss with rate $\gamma$, the vertices of the rings also feature an energy offset $\Delta$. The alternating hopping rates $t$ and $t_1$ along the lattice divide the system into unit cells (labeled by $m$) consisting of sublattice sites (labeled by $a$ and $b$).
Such a model, as we argue below, is analogous to the non-Hermitian SSH model studied in Ref.~\cite{SkinT2}. An outstanding feature of such a non-Hermitian model is the presence of non-Hermitian skin effects, where all eigenstates of the system under the open-boundary condition become localized at boundaries. The skin effects also give rise to the breakdown of the conventional bulk-boundary correspondence, a fundamentally important phenomenon unique to non-Hermitian topological systems which has stimulated intense theoretical and experimental studies~\cite{SkinT1,SkinT2,SkinT3,SkinT4,XueSkin,ThomaleSkin,CoulaisSkin}. Key to the non-Hermitian SSH model is the non-reciprocal hopping between two sublattice sites within a unit cell, which, in our case, is guaranteed by the non-reciprocal transport of the dissipative AB ring. A qualitative understanding of the proposed setup can be obtained in the weak-coupling limit ($t_0\ll \Delta,\gamma,t$), where the effective Hamiltonian reads (setting $\phi=\pi/2$)~\cite{sm}
\begin{align}
H_{\rm skin}&=\sum_m \big[\tilde{\Delta}(c^\dag_{m,a}c_{m,a}+c^\dag_{m,b}c_{m,b})+(t+\tilde{\gamma})c^\dag_{m,a}c_{m,b}\nonumber\\
&+(t-\tilde{\gamma})c^\dag_{m,b}c_{m,a}+t_1( c^\dag_{m+1,a}c_{m,b}+\text{H.c.})\big].\label{eq:skinH}
\end{align}
Here $c_{m,a(b)}$ ($c^\dag_{m,a(b)}$) is the annihilation (creation) operator for the $a(b)$ sublattice in the $m$th unit cell, the complex Stark shift $\tilde\Delta$ and the complex differential hopping rate $\tilde{\gamma}$ are functions of $t_0$, $\Delta$ and $\gamma$~\cite{sm}. Equation~(\ref{eq:skinH}) is similar to the non-Hermitian SSH model in Ref.~\cite{SkinT2}, only with additional complex energy-shift terms on the sublattice sites and a complex $\tilde{\gamma}$. We have numerically checked that, under the open-boundary condition, the model in Fig.~\ref{fig4} features non-Hermitian skin effects and the breakdown of conventional bulk-boundary correspondence, even for parameters beyond the weak-coupling regime~\cite{sm}. Building upon the experimental scheme reported here, the configuration in Fig.~\ref{fig4} can be readily implemented in cold atoms using a two-component BEC, with one component prepared in the momentum lattice, and the other subject to laser-induced loss~\cite{Li2019}. The intra- and inter-species hoppings can be induced by Bragg lasers or by microwave fields.

{\it Conclusion:---}
With highly tunable, non-reciprocal transport properties, dissipative AB rings are useful building blocks for applications in quantum simulation and quantum information. Our experiment therefore not only lays the groundwork for investigating non-reciprocal many-body dynamics in cold atoms, but also prepares for the simulation of intriguing non-Hermitian physics or the design of useful quantum device in the quantum many-body setting of cold atoms.

{\it Acknowledgement:---}
We thank Professor Youquan Li and Lihua Lv for helpful discussions. We acknowledges support from the National Key R$\&$D Program of China under Grant Nos. 2018YFA0307200, 2016YFA0301700 and 2017YFA0304100. National Natural Science Foundation of China under Grant Nos. 91636104, 11974331 and 91736209. Natural Science Foundation of Zhejiang province under Grant No. LZ18A040001, and the Fundamental Research Funds for the Central Universities. B.G. acknowledges support from the National Science Foundation under Grant No. 1707731.

\bibliographystyle{apsrev4-1}
\bibliography{flux_non_Hermitian_chain}

\clearpage
\begin{widetext}
\appendix

\renewcommand{\thesection}{\Alph{section}}
\renewcommand{\thefigure}{S\arabic{figure}}
\renewcommand{\thetable}{S\Roman{table}}
\setcounter{figure}{0}
\renewcommand{\theequation}{S\arabic{equation}}
\setcounter{equation}{0}

\newpage
\section{Supplemental Materials}
Here we provide more details on the experimental procedure, the theoretical characterization of an ideal dissipative AB ring, the derivation of the full and effective Hamiltonians, the impact of interactions, and an example for the application of dissipative AB rings in quantum simulation.

\section{Experimental preparation and detection}

After $^{87}$Rb atoms are collected in a magneto-optical trap (MOT), we compress the MOT and then apply optical molasses for $30$ms, which cools the atoms to a temperature of about $10\mu$K. At the same time, the dipole trap is turned on to directly load cold atoms into the dipole trap. The evaporative cooling is performed in the dipole trap for $18$s, which finally creates a BEC of $\sim 6\times10^4$ atoms.

To generate the momentum lattice, discrete momentum states are coupled with multi-frequency Bragg laser pairs, where different frequency components are imprinted with two acoustic optical modulators (AOMs). One AOM shifts the incoming beam by $-100$MHz, and the other shifts another $100$MHz$+\sum_n f_n$, where $f_n=(2n+1)\times4E_r/h$. In total, the two AOMs shift the frequency by $\sum_n f_n$, such that
momentum states $|n\rangle\leftrightarrow|n+1\rangle$ are resonantly coupled. In our experiment, we generate $15$ momentum-lattice sites by applying Bragg laser pairs with $14$ different frequencies.
The parameters $t$, $t'$, $t_c$ and $t_r$ can be independently tuned by adjusting the laser intensities for the corresponding frequency component. Since interaction effects are negligible in our experiment (see the section ``Impact of interaction'' in this Supplemental Materials for details), we treat BEC atoms as non-interacting particles when generating the momentum lattice.

After applying the Bragg beams, we turn off all the laser beams and let the atoms fall freely in space for $20$ms, before we take a camera image of the atoms, measuring the population in different momentum states.

\section{Ideal dissipative AB ring}
The Hamiltonian for an ideal dissipative AB ring shown in Fig.~\ref{fig1}(a) can be written as
\begin{align}\label{supp:eq1}
H_{\rm eff} = -i\gamma c_{-1}^\dagger c_{-1}^{} + \Big[\sum_{n\neq -1,0} t c_n^\dagger c_{n+1} + te^{-i\frac{\phi}{2}}c_{-1}^\dag c_0
+ te^{-i\frac{\phi}{2}}c_0^\dag c_1 + t'c_{1}^\dagger c_{-1}^{} + \text{H.c.}\Big],
\end{align}
where $c^\dagger_n$ ($c_n$) creates (annihilates) a particle at site $n$. This Hamiltonian is non-Hermitian, with broken inversion and time-reversal symmetries for finite $\gamma$ and $\phi$, respectively~\cite{Jin2018}.

For the simulation, we consider a $101$-site lattice, with open boundaries at sites $\pm 50$ and with the AB ring on sites $-1$, $0$ and $1$. We
initialize a single-particle state on the right or left end of the lattice ($\pm 50$) for the right- or left-moving transport, respectively, and numerically evolve the system under (\ref{supp:eq1}). Typical time-dependent population distributions are shown in Fig.~\ref{fig1}(b) of the main text, where we fix $\phi=3\pi/2$ and $\gamma/t=1$ and demonstrate the non-reciprocal transport of the dissipative AB ring.

In Figs.~\ref{fig:ideal}(a) and (b), we show the numerically calculated population loss as functions of phase $\phi$ and loss rate $\gamma$, respectively. Here the population loss is defined as the decrease of the population on the lattice, at the end of a time evolution with the duration $50\hbar/t$. The numerical results of the population loss in the ideal case are consistent with our experimental measurements.

\begin{figure}[]
\includegraphics[width=0.8\textwidth]{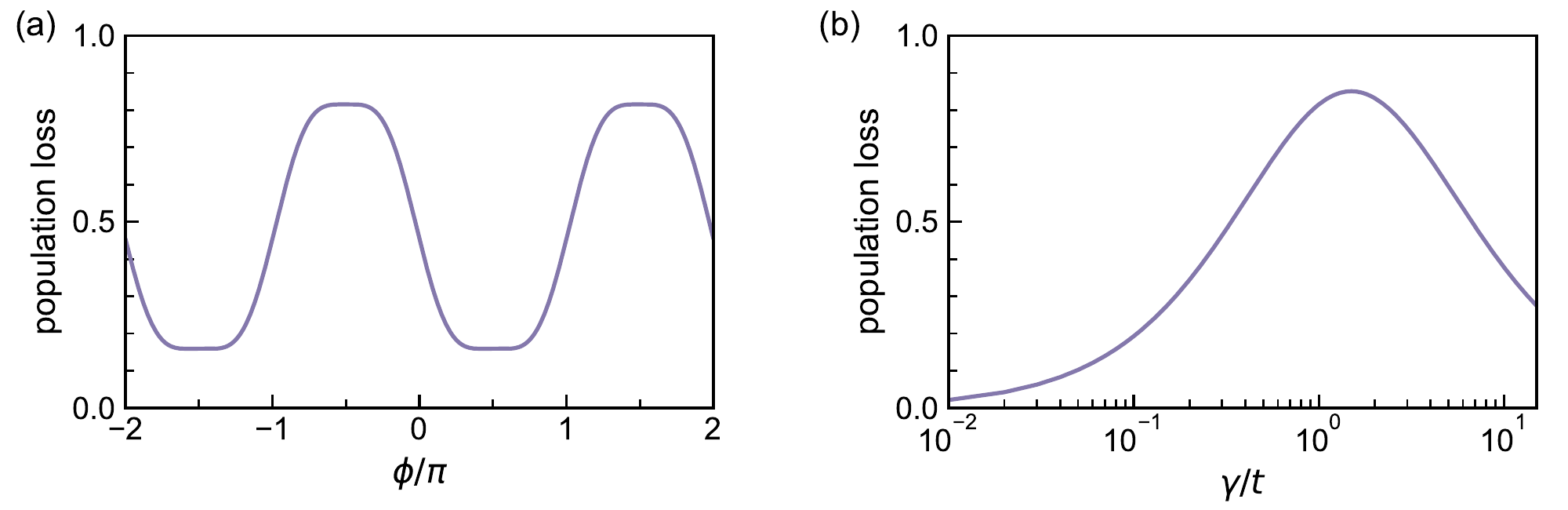}
\caption{Numerical simulation of transport properties of an ideal dissipative AB ring, calculated using Eq.~(\ref{supp:eq1}). (a) Dependence of the population loss on the synthetic magnetic flux $\phi$, where we fix $\gamma=t$ for the simulation. (b) Dependence of the population loss on the loss rate $\gamma$, where we fix $\phi=3\pi/2$. For both cases, we initialize the single-particle state at the lattice site $n=-50$ to the left of the ring, and let it evolve for a time of $50\hbar/t$.
\label{fig:ideal}}
\end{figure}

\section{Derivation of the full and effective Hamiltonians}

{
We start from a particle in a light field. Under the dipole approximation, the single-particle Hamiltonian can be written as
\begin{equation}
 H=\frac{{\hat{\bf p}}^2}{2M}+\hbar\omega_e\left| e \right\rangle \left\langle e \right|+\hbar {\omega _g}\left| g \right\rangle \left\langle g \right|+{\bf d}\cdot{\bf E}.
\end{equation}
Here $|g\rangle$ and $|e\rangle$ are the ground and excited states for the particle. The interaction between the particle and light is the dipole interaction ${\bf{d\cdot E}}$. The electric field ${\bf E}={{\bf E}_+}+{{\bf E}_-}$, and
\begin{equation}
{{\bf E}_+}={{\bf E}_+} \cos({{\bf k}_+}\cdot{\bf x}-{\omega_+}t+\tilde\phi_+),
\end{equation}
\begin{equation}
{{\bf E}_-}=\sum\limits_j {{\bf E}_j \cos({{\bf k}_j}\cdot{\bf x}-{\omega_i}t+\tilde\phi_j)}.
\end{equation}
Here ${{\bf E}_-}$ corresponds to the laser with multi-frequency components. We define ${\Omega_+}=\left\langle e \right| {\bf d}\cdot{\bf {E_+}}\left| g \right\rangle/\hbar$, ${\Omega_j}=\left\langle e \right| {\bf d}\cdot{\bf {E}}_j\left| g \right\rangle/\hbar$, and ${{\bf k}_+}=k\hat x$, ${{\bf k}_j} \simeq -k{\hat x},   \forall j$. We apply the rotating-wave approximation, adiabatically eliminate the excited state $|e\rangle$, and expand the external motion of the ground state in the discrete momentum lattice with $\left| \psi \right\rangle=\sum\limits_n { c_n}{e^{i2nkx}}\left| n \right\rangle \otimes \left| g \right\rangle$ ($n\in \mathbb{Z}$). Then we {obtain} a Hamiltonian in the interaction picture,
\begin{equation}
 H_{\rm full}=\sum_n \sum\limits_j \frac{\hbar\tilde\Omega_j}{2}e^{i(2n+1)\frac{4E_r}{\hbar}t} e^{-i(\omega_+-\omega_i)t}e^{i(\tilde\phi_+-\tilde\phi_j)} | n+1 \rangle \langle n | + \text{H.c.}.\label{eq:suppfullH}
\end{equation}
Here we define a two-photon Rabi {frequency}, $\tilde\Omega_i=\frac{\Omega_i\Omega_+}{2\Delta}$, and $E_r=\hbar^2 k^2/2m$ is the one-photon recoil energy. In our experiment, we let the laser frequencies in the right-hand beam are: $\omega_j = \omega_+ - 4(2j+1)E_r/\hbar$ ($j\in Z$) and an additional $\omega_- = \omega_+$; see Fig. 1(c).  With the choices of $\tilde\phi_j-\tilde\phi_+ = \phi_j$ and $\tilde\phi_--\tilde\phi_+=\phi_-$, we have
\begin{equation}
H_\text{full} =\sum\limits_n\left[ \frac{\hbar\tilde\Omega_-}{2}e^{i4(2n+1)E_rt/\hbar}e^{-i\phi_-}|n+1\rangle\langle n| + \sum_{j} \frac{\hbar\tilde\Omega_j}{2}e^{i8(n-j)E_rt/\hbar}e^{-i\phi_j}|n+1\rangle\langle n| \right] + \text{H.c.},
\end{equation}
which can be rewritten as
\begin{equation}
H_\text{full} = \sum_{\ell} \left( H^{[\pm2\ell]} e^{\pm i8\ell E_rt/\hbar} + H^{[\pm(2\ell+1)]} e^{\pm i4(2\ell+1) E_rt/\hbar} \right)
\end{equation}
with $\ell \in \mathbb{N}$. Here
\begin{eqnarray}
&H&^{[+2\ell]} = \sum_n \frac{\hbar\tilde\Omega_{n-\ell}}{2} e^{-i\phi_{n-\ell}} |n+1\rangle \langle n| + \sum_n\frac{\hbar\tilde\Omega_{n+\ell}}{2} e^{-i\phi_{n+\ell}} |n\rangle \langle n+1|, \\[0.5em]
&H&^{[+(2\ell+1)]}  =  \frac{\hbar\tilde\Omega_-}{2} \left(e^{-i\phi_-}|\ell+1\rangle\langle \ell| + e^{i\phi_-}|-\ell-1\rangle\langle -\ell|\right),
\end{eqnarray}
with $H^{[-2\ell]} = H^{[+2\ell]\dagger}$, $H^{[-(2\ell+1)]} = H^{[+(2\ell+1)]\dagger}$.
Apparently, all the resonant two-photon Bragg diffractions are described by the term
\begin{equation}
H^{[0]} = \sum_n \frac{\hbar\tilde\Omega_n}{2} e^{-i\phi_n} |n+1\rangle\langle n| + \text{H.c.}.
\end{equation}
$H^{[\pm(2\ell+1)]}$ describes the off-resonant couplings $|\ell\rangle\leftrightarrow|\ell+1\rangle$ and $|-\ell\rangle\leftrightarrow |-\ell-1\rangle$, induced by the Bragg lasers with frequencies $\{\omega_+, \omega_-\}$; while $H^{[\pm 2\ell]}$ describes the off-resonant couplings $|n\rangle\leftrightarrow|n+1\rangle$, which are induced by Bragg lasers pairs with frequencies
$\{\omega_+, \omega_{n-\ell}\}$ and $\{\omega_+,\omega_{n+\ell}\}$.

Following the treatment in~\cite{Giese2013}, we derive a time-independent, effective Hamiltonian up to the second-order corrections in the off-resonant couplings
\begin{equation}\label{eq6}
H_{\rm eff}= H^{[0]} + \sum\limits_{\ell\in N^+} \frac{[H^{[+2\ell]}, H^{[-2\ell]}]}{4(2\ell)E_r} + \sum_{\ell\in N}\frac{[H^{[+(2\ell+1)]}, H^{[-(2\ell+1)]}]}{4(2\ell+1)E_r}.
\end{equation}
Letting $t_j = \hbar\tilde\Omega_j/2$ and $t_- = \hbar\tilde\Omega_-/2$, we have
\begin{eqnarray}
[H^{[+2\ell]}, H^{[-2\ell]}] &=& \sum_n\left\{ \left[(t_{n-\ell}t_{n+\ell-1} e^{-i(\phi_{n-\ell}-\phi_{n+\ell-1})}- t_{n+\ell}t_{n-\ell-1}e^{-i(\phi_{n+\ell}+\phi_{n-\ell-1})}) |n-1\rangle\langle n+1|+ \text{H.c.}\right]\right.\nonumber \\[0.5em]
&+&\left. (t^2_{n+\ell}+t^2_{n-\ell-1}-t^2_{n-\ell}-t^2_{n+\ell-1}) |n\rangle\langle n| \right\}, \\[1em]
[H^{[+(2\ell+1)]}, H^{[-(2\ell+1)]}] &=& t_-^2(|\ell+1\rangle\langle \ell+1|+|-\ell-1\rangle\langle-\ell-1| - |\ell\rangle\langle\ell|-|-\ell\rangle\langle-\ell |) \nonumber\\[0.5em]
&+& t_-^2e^{-i2\phi_-}(|\ell+1\rangle\langle \ell|-\ell\rangle\langle-\ell-1|-|-\ell\rangle\langle-\ell-1|\ell+1\rangle\langle\ell|) \nonumber\\[0.5em]
&+& t_-^2e^{i2\phi_-}(|-\ell-1\rangle\langle-\ell|\ell\rangle\langle\ell+1| - |\ell\rangle\langle\ell+1|-\ell-1\rangle\langle-\ell|).\label{eq8}
\end{eqnarray}
Note that the last two terms in the right hand of Eq.~(\ref{eq8}) should vanish except for $\ell = 0$, which contribute to the synthetic magnetic flux in the AB ring. Then, the Hamiltonian (\ref{eq6}) can be expanded as
\begin{equation}\label{eq10}
H_{\rm eff} = \sum_n (t_n^{(1)}|n+1\rangle\langle n| + \text{H.c.}) + \sum_n (t_n^{(2)} |n+1\rangle\langle n-1|+\text{H.c.})  + \sum_n\Delta_n |n\rangle\langle n|,
\end{equation}
with $t_n^{(1)} = t_ne^{-i\phi_n}$, and
\begin{eqnarray}
t_n^{(2)} &=& \left\{ \begin{array}{lcl} \frac{t_-^2}{4E_r}e^{-i2\phi_-} + \sum\limits_{\ell\in N^{+}}\frac{1}{8\ell E_r}\left[t_{-\ell}t_{\ell-1} e^{-i(\phi_{-\ell}-\phi_{\ell-1})}- t_{\ell}t_{-\ell-1}e^{-i(\phi_\ell+\phi_{-\ell-1})}\right] & \mbox{for} & n = 0 \\[2em]
\sum\limits_{\ell\in N^{+}}\frac{1}{8\ell E_r}\left[t_{n-\ell}t_{n+\ell-1}e^{-i(\phi_{n-\ell}-\phi_{n+\ell-1})} - t_{n+\ell}t_{n-\ell-1}e^{-i(\phi_{n+\ell}+\phi_{n-\ell-1})}\right]  & \mbox{for} & n \neq 0 \end{array}\right. , \label{eq11}\\[2em]
\Delta_n & = &
\frac{ t_-^2}{4E_r}\left(\frac{1}{2|n|-1}-\frac{1}{2|n|+1}\right) + \sum\limits_{\ell\in N^{+}}\frac{1}{8\ell E_r}(t^2_{n+\ell}+t^2_{n-\ell-1}-t^2_{n-\ell}-t^2_{n+\ell-1}).
\end{eqnarray}

Setting $t' =  t_-^2/4E_r$, $\phi' = -2\phi_-$, and neglecting the terms with denominators larger than $8E_r$ ($\sim h\times16.2~\text{kHz}$ in our experiment), we then have

\begin{equation}\label{eq14}
H_{\text{eff}} = \sum_n t'\left(\frac{1}{2|n|-1}-\frac{1}{2|n|+1}\right)c_n^\dagger c_n+ \left(\sum_n t_n e^{i\phi_n} c_n^\dagger c_{n+1} + t'e^{i\phi'} c_{1}^\dagger c_{-1}^{} + \text{H.c.}\right).
\end{equation}
In the case of $\phi_{-1}=\phi_0=-\phi/2$, and setting $\phi'$ and all other $\phi_n$ to $0$, Eq.~(\ref{eq14}) reduces to the effective Hamiltonian (1) in the main text, except for the light-shift terms proportional to $t'$. While the light-shift terms are induced by the off-resonant $\omega_-$ frequency components, they can be compensated by introducing additional detuning $\delta_j$ to each frequency components, with $\omega_j = \omega_+ - \delta_j - 4(2j+1)E_r/\hbar$ and $\omega_-=\omega_+$. For such a purpose, we need
\begin{equation}
\delta_j = \Delta_{j+1} - \Delta_j,
\end{equation}
where $\Delta_j=t' \left[1/(2|j|-1)-1/(2|j|+1)\right]$. Experimentally, this is easily achieved by modulating the AOMs. Specifically, in our $15$-site experiment, with the parameters $t_n = h\times 1.25(2)$kHz, $t_- = h\times 3.20(5)$kHz, the compensation detunings for the adjacent-site couplings are calculated to be (in units of $h\times$kHz)
\begin{equation}
\{0.003,  0.006,  0.011,  0.025,  0.075,
        0.525, -2.627 ,  2.627 , -0.525, -0.075,
       -0.025, -0.011, -0.006, -0.003\}.
\end{equation}
In our experiment, we only compensate for the central four sites with the largest shifts.

With such a compensation scheme, Eq.~(\ref{eq14}) becomes
\begin{equation}
H_\text{eff} = \sum\limits_{n<-1} t_n (e^{i\phi_n}c_n^\dagger c_{n+1} + \text{H.c.}) + \left(\sum_{n\geq-1} t_ne^{i\phi_n}c_n^\dagger c_{n+1}^{} + t'e^{i\phi'}c_{1}^\dagger c_{-1}^{} + \text{H.c.}\right).
\end{equation}
For the experimentally relevant condition $t_{-2} = t_c$, $t_{n<-2}=t_r$, and $t_n=t$,  we apply the second-order perturbation to eliminate the reservoir (sites with $n<-1$), and derive the effective Hamiltonian as in the main text
\begin{align}
H_{\rm eff} = -i\gamma c_{-1}^\dagger c_{-1}^{} + \Big[\sum_{n\geq1} t c_n^\dagger c_{n+1} + te^{-i\frac{\phi}{2}}c_{-1}^\dag c_0+ te^{-i\frac{\phi}{2}}c_0^\dag c_1 + t'c_{1}^\dagger c_{-1}^{} + \text{H.c.}\Big],\label{supp:Heff}
\end{align}
where $\gamma= t_c^2/t_r$, and we have taken the phase convention: $\phi'=0$, $\phi_n=0$ except for $\phi_{-1}=\phi_0=-\phi/2$.  }

{
\section{Light-shift compensation and phase schemes}

Here we discuss the impact of light-shift compensation and phase schemes on the simulation of dissipative AB ring. We discuss two types of phase schemes: phase scheme $1$, with
 $\phi_- = 0$ and $\phi_0 = \phi_{-1} = -\phi/2$, is the one we adopt in the main text; phase scheme $2$, with $\phi_0 = \phi_{-1} = 0$ and $\phi_- = \phi/2$. Physically, under phase scheme $1$, the synthetic magnetic flux is induced by the phase of coupling $|-1\rangle\leftrightarrow|1\rangle$, while under phase scheme $2$, the flux is induced by the a symmetric distribution of the phases on $|0\rangle\leftrightarrow|-1\rangle$ and $|1\rangle\leftrightarrow|0\rangle$.

In Fig.~\ref{figS2}(a), we show the impact of phase schemes on the non-reciprocal transport.
Whereas numerical results from the effective Hamiltonian Eq.~(\ref{supp:Heff}) do not depend on the phase schemes, for simulations using the full Hamiltonian Eq.~(\ref{eq:suppfullH}), as well as for our experimental measurements, the choice of phase scheme does affects the non-reciprocal transport. In particular, phase scheme $1$ offers a better simulation of the ideal dissipative AB ring than scheme $2$.

In Fig.~\ref{figS2}(b), we show the impact of light-shift compensation. Here, we adopt the phase scheme $2$, which does not affect the overall conclusion. It is apparent that the light-shift compensation has a more significant impact on the transport than the choice of phase scheme, and the compensation enables a better simulation of the ideal dissipative AB ring, which is modeled by the effective Hamiltonian Eq.~(\ref{supp:Heff}).

\begin{figure*}[tbp]
\includegraphics[width=0.8\textwidth]{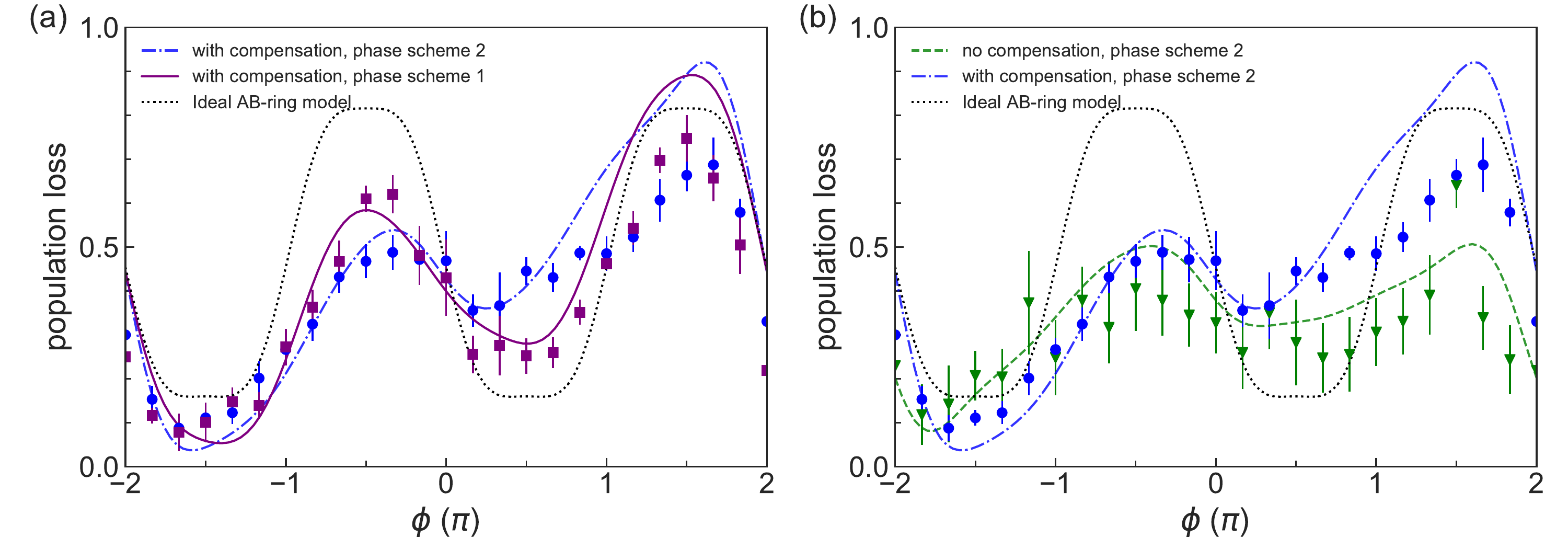}
\caption{\label{figS2} The population loss $\mathcal{P}_\ell$ as a function of $\phi$ under different configurations, after an evolution time of $3\hbar/t$. For both panels, the black dotted curve is numerical simulation under the ideal dissipative AB-ring model Eq.~(\ref{supp:Heff}).
(a) Impact of phase scheme. The blue dash-dotted curve is the numerical calculation under phase scheme 2, with $\phi_0 = \phi_{-1} = 0$ and $\phi_- = \phi/2$. The blue dots with error bars are the experimental measurement. The purple solid curve is numerical simulation with light-shift compensation under phase scheme $1$, with $\phi_- = 0$ and $\phi_0 = \phi_{-1} = -\phi/2$. The purple dots with error bars are the corresponding experimental measurement.
(b) Impact of compensation. We adopt the phase scheme $2$. The green dashed curve is numerical simulation without light-shift compensation, and the green dots with error bars are the experimental measurement under the same conditions. In all cases, we fix the effective loss rate $\gamma/t=1$, and initialize the BEC on site $|0\rangle$.
}
\end{figure*}

\section{Impacts of interaction}

Let's discuss the impact of interaction on the non-reciprocal transport now.
We consider a homogeneous BEC with density $\rho$. Under the Hartree-Fock approximation, the dynamics of a interacting, dissipative AB ring is governed by the equation
\begin{equation}\label{eqU}
i\hbar\partial_t \Psi = \tilde{H}\Psi,
\end{equation}
where $\Psi = [\psi_{-1},\psi_0,..., \psi_n]^{T}$, with $\psi_n$ the mean-field wavefunction for site $|n\rangle$. The matrix elements of $\tilde{H}$ are
\begin{eqnarray}
\tilde{H}_{m,n} &=& H_\text{eff}^{m,n} ~\text{for} ~ m\neq n \nonumber\\
\tilde{H}_{n,n} & = & H_\text{eff}^{n,n} + U[|\psi_n|^2+2\sum\limits_{m\neq n}|\psi_m|^2],
\end{eqnarray}
where $U = 4\pi\hbar^2 a_s \rho/\mu$, $a_s$ is the $s$-wave scattering length, and $\mu$ is the atomic mass. $H_\text{eff}^{m,n}$ is the matrix element of the Hamiltonian Eq.~(\ref{supp:Heff}) under the basis $\{|n\rangle\}$.

We perform simulations using Eq.(\ref{eqU}) under various parameters for a finite-size lattice with $n=7$. As shown in Fig.~\ref{figS3}(a)(b),
regardless of the value of $\phi$, transmission dominates for small $U/t$, whereas atoms become localized at the initial state $|0\rangle$ for large enough $U/t$ ($U/t>6$ from our calculation). Such an interaction-induced localization originates from an extra exchange term for interacting atoms in different momentum states~\cite{An2018}. In  Fig.~\ref{figS3}(c), we show the population loss $\mathcal{P}_{\ell}$ as a function of $U/t$, after an evolution time of $3\hbar/t$.
Whereas the total loss does not change significantly for $U/t<4$, the transport is mostly blocked for $U/t>8$, due to the interaction-induced localization.

In our experiment, however, the interaction strength $U$ is estimated to be $\sim h\times 0.6~\text{kHz}$, while we typically have $t\sim h\times 1.25~\text{kHz}$, leading to $U/t\approx 0.5$. We thus conclude that interactions do not play a significant role on the transport behaviour in our experiment. Here the interaction strength $U$ is estimated using the parameters $a_s\approx 100a_0$ and $\rho\approx 7\times 10^{13}$cm$^{-3}$, where the BEC density is estimated using the Thomas-Fermi approximation.

\begin{figure}[b]
\includegraphics[width=0.8\textwidth]{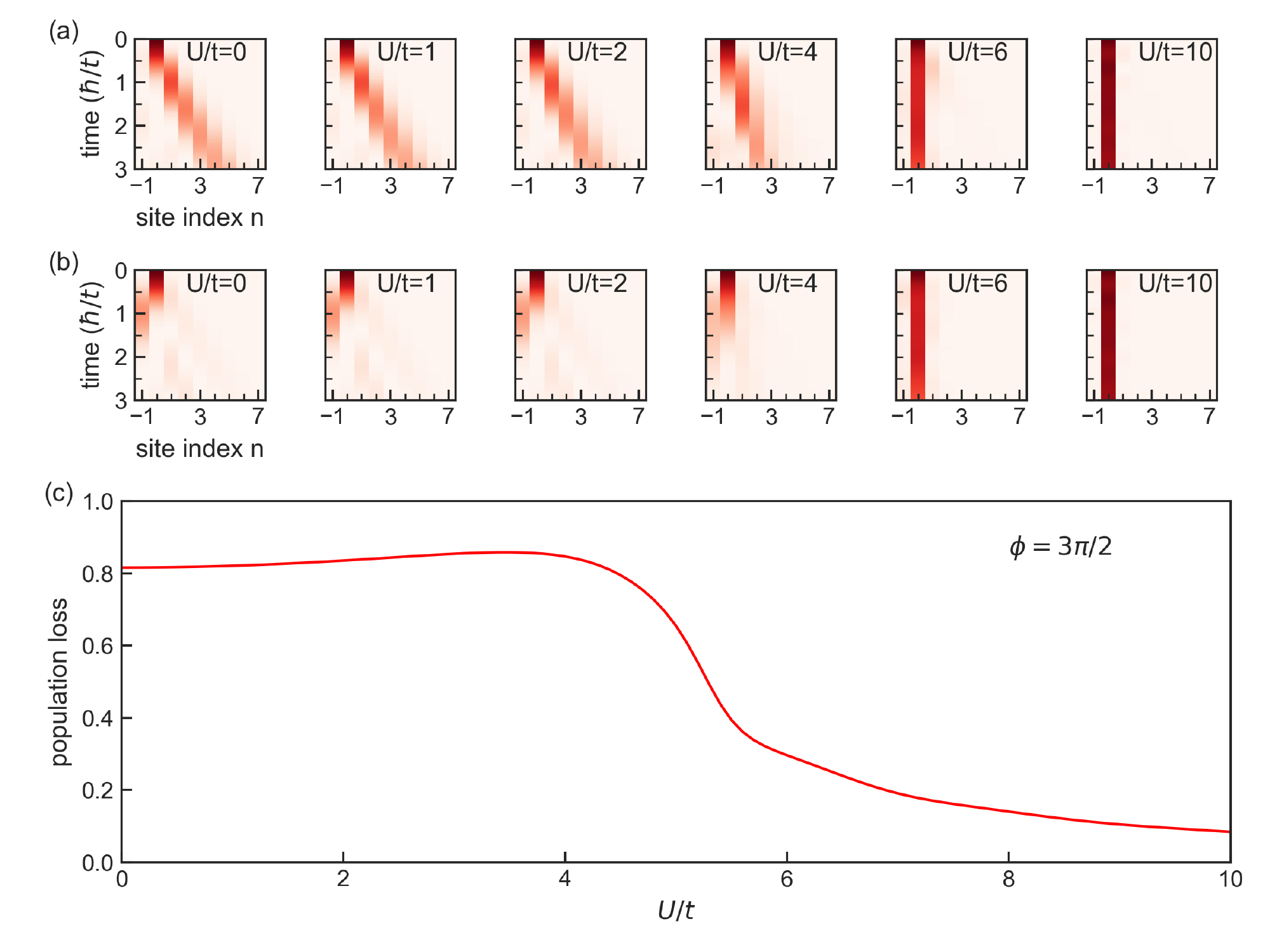}
\caption{\label{figS3} Transport in an interacting, dissipative AB ring. The simulations are performed with the effective Hamiltonian (\ref{supp:Heff}) for a finite lattice with $n=7$. Here we focus on two cases: (a) $\phi = \pi/2$ and (b) $\phi = 3\pi/2$, with $t'/t = 1$ and $\gamma/t = 1$. The time evolutions of atom distribution under different interaction strengths are shown. (c) Population loss as a function of the interaction strength.
}
\end{figure}
}

\begin{figure}[tbp]
\includegraphics[width=0.8\textwidth]{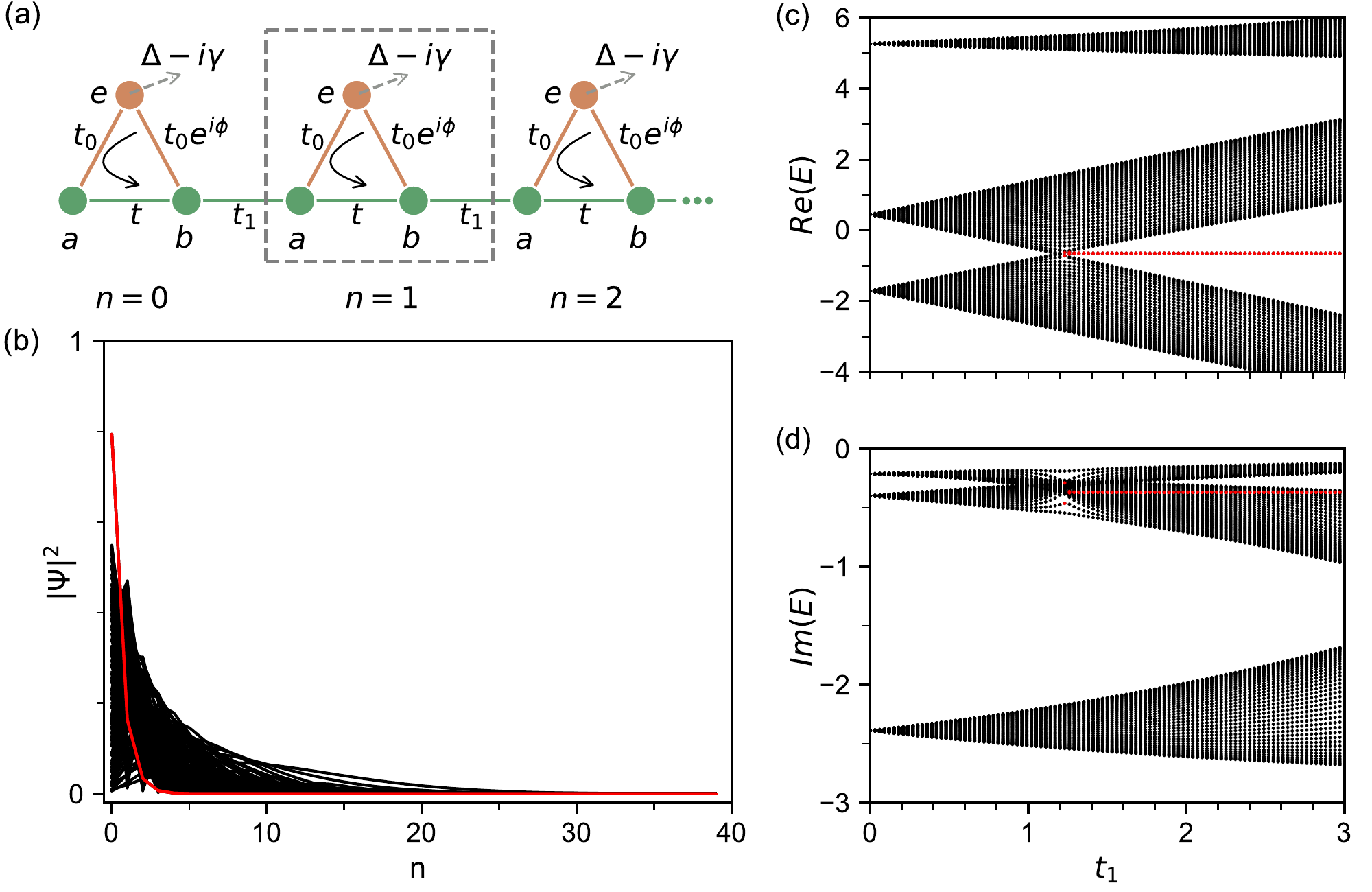}
\caption{Building topological lattice with non-Hermitian skin effect and non-Hermitian bulk-boundary correspondence. (a) Schematics for building non-Hermitian topological lattice with dissipative AB ring. See text for definition of parameters. (b) Eignestates wavefunctions under the open boundary condition demonstrate non-Hermitian skin effects. (c) Real and (d) imaginary eigenenergies of the system under the open boundary condition. In (b)(c)(d), edge states are marked in red, and bulks states are in black. We take $\Delta/t=4$, $\gamma/t=3$, $\phi=\pi/2$, and $t_0/t=2$ for calculations in (b), and further take $t_1/t=2$ in (c)(d).
\label{fig:figskin}}
\end{figure}

\section{Dissipative AB ring as key element for building non-Hermitian topological lattice}

As an example for its application, we explicitly demonstrate the utility of dissipative AB rings in studying non-Hermitian topology. Such a capability directly derives from the inherent non-reciprocity that we observe experimentally.

As illustrated in Fig.~\ref{fig:figskin}, we propose to assemble AB rings into a one-dimensional lattice, with each unit cell containing one such ring. Here $\Delta$ is the energy offset on site $e$ and $\gamma$ is the loss rate, $t_0$ is the coupling between sites $a$ and $e$, $t$ and $t_1$ are the hopping rates between $a$ and $b$ sites along the lattice. In such a model, the interplay of synthetic magnetic flux within the ring and the on-site dissipation on site $e$ leads to asymmetric hopping between sites $a$ and $b$ within each unit cell. The resulting lattice is then similar to the non-Hermitian Su-Schieffer-Heeger (SSH) model studied in Refs.~\cite{SkinT2}, which feature the fascinating non-Hermitian skin effects and the breakdown of conventional bulk-boundary correspondence. In fact, when site $e$ is weakly coupled to $a$ and $b$, the lattice model can be simplified, through second-order perturbation, to (we take $\phi=\pi/2$ for concreteness)
\begin{align}
H_{\text{skin}}&=\sum_n\left[
(-\frac{\Delta t_0^2}{\Delta^2+\gamma^2}-i\frac{\gamma t_0^2}{\Delta^2+\gamma^2})(c^\dag_{n,a}c_{n,a}+c^\dag_{n,b}c_{n,b})+t_1(c^\dag_{n+1,a}c_{n,b}+\text{H.c.})
\right.\nonumber\\
&\left.+(t+\frac{\gamma t_0^2}{\Delta^2+\gamma^2}-i\frac{\Delta t_0^2}{\Delta^2+\gamma^2})c^\dag_{n,a}c_{n,b}+(t-\frac{\gamma t_0^2}{\Delta^2+\gamma^2}+i\frac{\Delta t_0^2}{\Delta^2+\gamma^2})c^\dag_{n,b}c_{n,a}
\right],\label{eq:Hskin}
\end{align}
where $c_{n,a(b)}$ is the annihilation operator of site $a$ ($b$) in the $n$th unit cell. Apparently, Eq.~(\ref{eq:Hskin}) can be directly mapped to the non-Hermitian SSH model in Ref.~\cite{SkinT2}. Following the analysis in Ref.~\cite{SkinT2}, we find the gap-closing condition to be
\begin{align}
|t-\frac{\gamma t_0^2}{\Delta^2+\gamma^2}+i\frac{\Delta t_0^2}{\Delta^2+\gamma^2}||t+\frac{\gamma t_0^2}{\Delta^2+\gamma^2}-i\frac{\Delta t_0^2}{\Delta^2+\gamma^2}|=|t_1|^2,
\end{align}
which is different from that of a homogeneous system with periodic boundary condition. This shows the breakdown of conventional bulk-boundary correspondence.

While Eq.~(\ref{eq:Hskin}) is derived perturbatively, with the assumption that $t_0^2\ll \Delta^2+\gamma^2$, non-Hermitian skin effects and non-Hermitian bulk-boundary correspondence also exist in the full model corresponding to Fig.~\ref{fig:figskin}(a). This is demonstrated explicitly in Figs.~\ref{fig:figskin}(b)(c)(d). Note that topological edge states (red) in Figs.~\ref{fig:figskin}(c)(d) are not located at zero energy, due to a global complex energy shift [see the first term on the right of Eq.~(\ref{eq:Hskin})].

Experimentally, the lattice of coupled dissipative AB rings can be implemented along a momentum lattice and by using hyperfine-spin degrees of freedom of the atoms. The on-site loss $\gamma$ can be implemented using laser-induced dissipation through an electronically excited state.

\end{widetext}

\end{document}